\documentclass[preprint]{aastex}
\received{99/99/9999}
\revised{99/99/9999}
\accepted{99/99/9999}
\ccc{code}
\cpright{type}{yyyy}
\shorttitle{Radiative Recombination Edge in GRB~970828}
\shortauthors{Yoshida et al.}
\newcommand{\IAUC}{IAU Circ.}

\newcommand{\etal}{et al.}
\newcommand{\asca}{ASCA}
\newcommand{\sax}{BeppoSAX}

\newcommand{\rxte}{RXTE}
\newcommand{\asm}{ASM/RXTE}
\newcommand{\batse}{BATSE/CGRO}
\newcommand{\ipn}{IPN}

\slugcomment{\it Accepted for publication in the Astrophysical Journal Letters}

\begin{document}

\title{
A POSSIBLE EMISSION FEATURE IN AN X-RAY AFTERGLOW OF GRB~970828
AS A RADIATIVE RECOMBINATION EDGE
}

\author{
A.~Yoshida\altaffilmark{1,2},
M.~Namiki\altaffilmark{1},
D.~Yonetoku\altaffilmark{3,4},
T.~Murakami\altaffilmark{3,4},
C.~Otani\altaffilmark{1}, N.~Kawai\altaffilmark{1,4},
Y.~Ueda\altaffilmark{3}, R.~Shibata\altaffilmark{3}
and S.~Uno\altaffilmark{5}
}

\altaffiltext{1}{The Institute of Physical and Chemical Research,
2-1, Hirosawa, Wako, Saitama, 351-0198 Japan
}

\altaffiltext{2}{Department of Physics, Aoyama Gakuin University,
6-16-1, Chitosedai, Setagaya, Tokyo, 157-8572  Japan
}

\altaffiltext{3}{
Institute of Space and Astronautical Science,
3-1-1, Yoshinodai, Sagamihara, Kanagawa 229-8510 Japan
}

\altaffiltext{4}{
Department of Physics, Tokyo Institute of Technology,
2-12-1, Ookayama, Meguro, Tokyo 152-0033 Japan
}

\altaffiltext{5}{
Nihon Fukushi University, Faculty of Social and Information Sciences,
26-2 Higashihaemi-cho, Handa, Aichi 475-0012 Japan
}

\authoremail{ayoshida@phys.aoyama.ac.jp}

\begin{abstract}
    A gamma-ray burst of 28 August 1997 was localized by the All-Sky
Monitor on the Rossi XTE satellite and its coordinates were promptly
disseminated. An \asca\/ followup started 1.17 days after the burst as
a Target of Opportunity Observation and detected an X-ray
afterglow. The spectral data displayed a {\it hump} around $\sim
5~{\rm keV}$ and an absorption column of $7.1 \times 10^{21}~{\rm
cm^{-2}}$.  This {\it hump} structure is likely a recombination edge
of iron in the vicinity of the source, taking account of the redshift
$z = 0.9578$ found for the likely host galaxy of the associated radio
flare. Radiative Recombination edge and Continuum model can interpret
the spectrum from highly ionized plasma in a non equilibrium
ionization state.
The absorption could be also due to the medium presumably in the
vicinity of the GRB.
\end{abstract}


%%%%%%%%%%%%%%%%%%%%%%%%%%%%%%%%%%%%%%%%%%%%%%%%%%%%%%%%%%%%%%%%%%%%%%%%
\section{Introduction}

   The site and the distance scale of $\gamma$-ray bursts (GRBs) is
becoming clear initiated by the observational breakthroughs made for
their afterglows at longer wave lengths \cite{Costa97, JVP97, Bond97,
Frail97, Sahu97, Metzger97}, together with extensive
theoretical works \cite{ Pac86, Woo93, RM92, MR97,
Wijers97}.
We have learned so far that GRBs are one of the most distant and violent
astrophysical phenomena in the universe from many efforts in
multi-wavelengths study from radio to $\gamma$-ray \cite{Kul98}.

   However, the progenitors and site of GRBs are still unknown; although
the merging neutron star binary made in the baryon clean environment
was a popular scenario at the time when GRB afterglows are discovered,
recent observations appear to suggest more ``dirty'' circumstances
expected for hypernova \cite{Pac98}, collapsar \cite{Woo93}, or
supranova scenarios \cite{Vie98}.
Optical Transients (OTs) less GRBs (``optically dark'' GRBs), a major
part in known GRB samples, are suggested by several authors to be
embedded inside the dusty environment in host galaxies, which may be
star-forming region.

The key observation to investigate the GRB environment is maybe strong
iron features reported in X-ray afterglows \cite{Piro99, Ay99, Piro00,
Antonelli00}.  Especially Piro \etal\/ (2000) report an striking
results from the Chandra for GRB~991216.  It clearly revealed both an
iron K$\alpha$ emission line and a recombination edge interpreted to
be from highly ionized iron.

  A GRB of 28.73931 August 1997 (GRB~970828) is one of important events
so far because it was an ``optically dark'' GRB and showed a spectral
feature possibly from highly ionized iron together with a large
absorption column.
It was detected and quickly localized into a small region with $2'
\times 5'$ accuracy by the All-Sky Monitor (ASM) aboard the Rossi XTE
satellite (RXTE) \cite{Remi97, Smith97}.  Subsequently PCA/RXTE found
an X-ray afterglow of 0.5 mCrab (2--10 keV) about four hours after the
burst detection \cite{Marshall97} and ASCA detected a fading X-ray
source by a prompt follow-up and ROSAT gave more accurate location of
10 arcsec at a later epoch \cite{Marshall97, Mura97b, Greiner97}.
Although the X-ray afterglow was rather strong, a corresponding OT was
invisible down to $R \approx 23.8$ \cite{Groot98}. However a series of
VLA observations detected a weak (4.5 $\sigma$) radio flare at 8.46
GHz at the period 3.5 days after the GRB inside the ROSAT error
region. The subsequent optical observation by the Palomar 200-inch
telescope revealed extended sources around the radio position
\cite{Djor01}.  It shows that the radio source is located between two
bright optical peaks; Djorgovski \etal\/ (2001) suggest that the GRB
occured inside a dust lane (corresponding to a dark gap between two
bright peaks ``A'' and ``B'' in their paper), or at an interface of
two interacting galaxies.
Following spectroscopic study by the Keck on the galaxy ``A''
clearly displayed two emission lines which are interpreted as [O II] and
[Ne III] lines. This gives a redshift of $z = 0.9578$ for the galaxy
which is likely a host of GRB~970828 \cite{Djor01}.

In this paper we report on properties of an X-ray afterglow of this
interesting burst, GRB~970828, with the data obtained by \asca\/ and
discuss an emission {\it hump} and low-energy absorption seen in its
spectra.

%%%%%%%%%%%%%%%%%%%%%%%%%%%%%%%%%%%%%%%%%%%%%%%%%%%%%%%%%%%%%%%%%%%%%%%%
\section{ASCA Observation and Properties of the X-ray Afterglow}
   Because of the prompt dissemination of the ASM position and the
effort for a very quick operations of the satellite, A Target of
Opportunity Observation (TOO) by \asca\/ in the 0.5--10\,keV range was
conducted soon after the alerts based on the \asm\/ detection.
The observation with net exposure time of $\sim36~{\rm ks}$ was performed
during a period of August 29.91 -- August 30.85 UT beginning at 1.17
days after the GRB. Both two scientific instruments on \asca\/, the
SIS and the GIS detectors, saw an X-ray source of an average
2--10 keV flux of $F_{X} \sim 4 \times 10^{-13}~{\rm
erg~cm^{-2}s^{-1}}$ within the combined source error region given by
\asm\/ and \ipn\/ \cite{Smith97, Hurley97, Mura97b}.  The position of
the X-ray source was determined at R.A.= $18^h08^m32^s.2$ and
Decl.=$+59^{\circ}18'54''$ (J2000) with a 90 \% error radius of $0'.5$
by an image analysis on the datasets.

  The X-ray source clearly displayed a fading behavior during the
observation.  Estimated 2--10 keV fluxes are plotted in
Fig.\ref{fig:GRB970828_lc} together with that measured by PCA/RXTE in
the earlier epoch \cite{Marshall97}. We assume a power-law spectrum
with the photon index of $-2$ for this estimation. The source faded
from $\sim 1 \times 10^{-11}~{\rm ergs~cm^{-2}s^{-1}}$ (the PCA
observation at $t \sim 1.4 \times 10^4~{\rm s}$ to $\sim 3 \times
10^{-13}~{\rm ergs~cm^{-2}s^{-1}}$ (at the end of the \asca\/
observation at $t \sim 1.7 \times 10^{5}~{\rm s}$) where $t$ is an
elapsed time since the burst. Assuming a power-law fading model, the
decay of the X-ray afterglow can be represented by $t^{-1.44 \pm
0.07}$. The solid line in Fig.\ref{fig:GRB970828_lc} represents this
average fading behavior. Note that the error corresponds to 90 \%
confidence throughout this paper.

\subsection{Variability}

  Zooming in the \asca\/ data, the X-ray flux does not follow a simple
decay. A peak-like structure appears around $t = 1.25 \times
10^{5}~{\rm s}$.  This structure can be represented with a
Gaussian-line shape superposed on an overall decaying trend
represented by $t^{-1.44}$. From the fit $\chi^2$ turns out to be
$9.82$ with the degree-of-freedom (d.o.f.) of 14, while the simple
power-law gives $\chi^2=21.5$ ($17$ d.o.f.). Thus this variability is
significant with 99 \% confidence by an F-test for three additional
parameters (i.e., centroid, width and normalization of a line).

The BeppoSAX also saw a variability in the afterglow of GRB~970508
\cite{Piro97}. It is a rather broad ($\delta t > 10^{5}~{\rm s}$ at $t
\approx 10^{5}~{\rm s}$) ``bursting activity'', while ASCA's ``flare
activity'' is narrow in time; $\delta t/t \sim 0.05$.

\subsection{Spectral Structure}

We made a spectral
study for the periods which are indicated by lines labeled ``A'',
``B'' and ``C'' in Fig.\ref{fig:GRB970828_lc}.
In this analysis, data were jointly fitted with a single model
for those from four scientific detectors, SIS-0, SIS-1, GIS-2 and
GIS-3, carried by the ASCA.
There is an excess feature found around $4.8~{\rm keV}$ in the spectrum of
``B'', which looks like a {\it hump} or  a {\it broad line} over
the continuum.
An interpretation of this {\it hump} as an Fe K$_\alpha$ line
was presented in the previous paper \cite{Ay99}.
However an inferred redshift of $\sim 0.33$ did not
match to that of the host galaxy reported by Djorgovski et al. (2001).
We apply Radiative Recombination edge and Continuum (RRC) structure
for this feature as an alternative interpretation.

The RRC model is described as
\begin{eqnarray*}
\frac{d N}{d E} \propto
\left\{
\begin{array}{ll}
(kT_{e})^{-3/2} \exp(-E/kT_{e}) &  \mbox{if $E \ge 9.28~{\rm keV}$ ;}\\
0 &  \mbox{otherwise.}\\
\end{array}
\right.
\end{eqnarray*}
where $k$ is Boltzmann constant and
$T_{e}$ is an electron temperature of the plasma,
which can be determined from a width of a RRC structure.

Introducing three parameters for RRC, i.e.,
$E_{\rm edge}$, $kT_{e}$ and normalization,
we achieve a significant reduction of $\chi^2$, $\Delta \chi^{2} = 13.2$.
Hence a confidence level of the RRC model is found by an F-test to be
$99.3$ \%.
The best-fit gives
the edge energy of $E_{\rm edge} = 4.76^{+0.19}_{-0.25}~{\rm keV}$,
the electron temperature at the rest frame
$k T_{e} = 0.8^{+1.0}_{-0.2}~{\rm keV}$
and the integrated RRC flux of
$F_{\rm RRC} = 1.7^{+6.4}_{-1.3} \times 10^{-5}~{\rm
photons~cm^{-2}s^{-1}}$.
The 90 \% error range of $E_{\rm edge}$
above is consistent with the expected edge energy of H-like iron at
 $4.74~{\rm keV}$ and He-like iron at $4.51~{\rm keV}$ with a redshift
of $z = 0.9578$.
Thus the spectral feature
could be explained as a radiative recombination edge with a hard tail
(continuum) above the edge energy in highly ionized plasma.

It should be notable that
there is no redshifted iron emission line seen at the expected
energy for such ionized plasma.
We searched for a H-like iron emission line ($6.97~{\rm
keV}$ at the rest frame) in the spectrum at ``B''. As shown in
figure~\ref{fig:GRB970828_spec}, we found no emission line around
$6.97/(1+z) = 3.56~\rm{keV}$
with the upper limit of $F_{\rm line} \leq 1.5 \times 10^{-6}~{\rm
photons~cm^{-2}s^{-1}}$.
No significant line structure was found in ``A'' nor ``C''.
The best-fit parameters and
upper-limits on iron emission line and the RRC intensities are
summarized in table~\ref{tab:fit_res} and
table~\ref{tab:GRB970828_Fe}.

  From these fits, a significat absorption column density is 
found only in the spectrum of the period ``C'', 
of which $N_{\rm H}$ is evaluated to be
$7.08^{+0.32}_{-0.27} \times 10^{21}~{\rm cm^{-2}}$ at the observer
frame.  We have a significant reduction of $\Delta \chi^{2} = 20.5
 (84~{\rm d.o.f.})$ by adding an absorption, which
corresponds to a $4.3~\sigma$ confidence level.  This is significantly
larger than the galactic value towards the source direction calculated
to be $3.4 \times 10^{20}~{\rm cm^{-2}}$ by COLDEN at The Einstein
On-Line Service, Smithsonian Astrophysical Observatory.  The fitting
results are summarized in Table~\ref{tab:fit_res}.

\section{Discussions}

  In the previous paper (Yoshida \etal\/ 1999), we proposed that the
{\it hump} was likely corresponding to a red shifted Fe K$\alpha$
line.  Taking an assumption of the line at $6.7~{\rm keV}$ at the
source (i.e., He-like iron), the redshift was estimated to be
$z \sim 0.33 $.
Piro et al. (1998) reports a similar emission line feature
from the \sax\/ observations of the GRB~970508 afterglow \cite{Piro99}.
They imply that it is possibly an Fe line at the redshift $z = 0.835$
which was derived from the earlier optical work for GRB~970508
\cite{Metzger97}.
In the case of GRB~970828, on the contrary, an optical transient
corresponding to the GRB was invisible down to $R = 23.8$
\cite{Groot98} and one could not know its redshift from an OT.

However, recently published work on a radio transient and
a possible host galaxy of GRB~970828 shows that the burst source
likely sits in a distant galaxy of $z = 0.9578$. This is inconsistent
with the interpretation of the X-ray spectral structure as an Fe
K$\alpha$ line.
An alternative interpretation is that the structure could be due to
an iron recombination edge instead of a fluorescence line.
For instance, the Chandra observation of GRB~991216 \cite{Piro00}
has clearly shown a recombination edge
with rest energy of $9.28~{\rm keV}$ (H-like iron) as well as a
K$\alpha$ line at $z = 1.020$ which is consistent with the optical
spectroscopy reported by Vreeswijk et al. (2000).

There is, however, a sharp difference between the Chandra spectra and
the one of ASCA of GRB~970828. The Chandra has seen also an Fe
K$\alpha$ line together with a recombination edge (Piro \etal\/ 2000).
However there appears no significant emission feature around the
expected energies for the case of ASCA.  Inclusion of an additional
emission line into a fitting model does not improve fitting at all,
and gives an upper bound in a line intensity of $< 1.5
\times10^{-6}~{\rm photons~cm^{-2}s^{-1}}$ (see
Table~\ref{tab:GRB970828_Fe}).  This was puzzling for the ASCA
spectra. But it could be understood in the case that plasma in
non-equilibrium ionization states are responsible for emission of the
RRC and the line. A scenario for emission mechanism of such an
apparently line-less recombination edge is described and discussed in the
companion paper (Yonetoku \etal\/ 2001).

A large absorption column is the other puzzle in the X-ray afterglow.
Taking $z=0.9578$ for the GRB source, the intrinsic column density of
the period ``C'' is derived by the fitting with the model, $\exp( -
N_{\rm H}^{\rm gal} \sigma(E) ) \ast \exp( -N_{\rm H}^{\rm src}
\sigma((1+z)E) ) \ast E^{-\alpha}$, to be $N_{\rm H}^{\rm
src}=(3.13^{+1.76}_{-1.29}) \times 10^{22}~{\rm cm^{-2}}$ with the
galactic density $N_{\rm H}^{\rm gal} = 3.4 \times 10^{20}~{\rm
cm^{-2}}$.  Although this column density coupled with the observed
change of power-law index, the large absorptions were measured in
several other X-ray afterglows; even the famous ``optically bright'',
GRB~990123 \cite{Yone00}.

No optical afterglow was discovered for GRB~970828 down to $R = 23.8$
\cite{Groot98}. The R band extinction should be large
from the above column density.  This would imply that the matter is 
near the
source in its host galaxy.
Djorgovski \etal\/ (2001) suggest that the GRB source sits inside
dusty environment of the host galaxy from the observations of the radio
flare and the optical images for its coordinate.

It may be also notable that the intrinsic absorption column is apparently 
variable.  An absorption is not necessary for the pre-flare and
the in-flare spectra (``A'' and ``B'') with 90\% upper-bounds 
of  $3.65 \times 10^{21}$ and 
$5.54 \times 10^{21}~{\rm cm^{-2}}$ respectively, 
while a large $N_{\rm H}$ is required for the post-flare period ``C'';
$7.08^{+3.21}_{-2.75} \times 10^{21}~{\rm cm^{-2}}$. 
Although the variability is marginal and the absorption column is 
coupled with the continuum, 
this implies that the medium which absorbed the X-ray 
would have been changed during the periods, ``A'' to ``C''.
Some authors have studied on time-dependent absorption
in the GRB environments \cite{PL98,Bot99}.
Perna \& Loeb (1998), for instance, suggest that 
the afterglow radiation ionizes the matter around the source 
and hence the absorption should 
decay in the later epoch instead of {\it increasing}, 
considering H\,I cloud with the density of 100\,cm$^{-3}$ 
within $\sim1$\,pc from the source. 
The inferred density from the current observation 
is much larger, $\sim10^{22}$\,cm$^{-2}$,
and could imply much denser medium.
The blast wave moving nearly with the light speed 
would have been at $\sim4\times10^{15}$\,cm 
from the source at the period ``C'' ($(1.4-1.7)\times10^5$\,s from the burst).
The size of the observable {\it beam} of afterglow radiation increases 
$\propto 1/\Gamma$
as the bulk Lorentz factor $\Gamma$ decreases.
If a dense matter was in the vicinity near the edge of the {\it beam},
it might have caused an increase of absorption 
as observed.
The {\it variable} absorption might be an indication 
for a strong inhomogeneous matter distribution 
in the vicinity of the GRB source. 
This is likely 
apart from a dust lane in a larger scale, which could have obscured
an optical transient \cite{Djor01}.

\acknowledgments

We are grateful to George Djorgovski and Shri Kulkarni for discussions
of the radio and optical observations.  We would like to thank Kuniaki
Masai and Masaru~Matsuoka for
valuable discussions for Fe emission mechanism.
We acknowledge all the \asca\/ team members, especially the operation
team, to allow us to make a quick TOO of this source, and express our
gratitude to the \asm\/ team for the prompt information on the GRB
location.  This work is indebted to a large international collaboration;
the authors are grateful to to \rxte, \ipn and \batse\/  teams for
their efforts. We thank the anonymous referee for the helpful comments.

\clearpage

\figcaption[]{
The X-ray light curve obtained with \asca\/ is plotted together
with the flux reported by Marshall et al (1997).
Here we assume a power-law spectrum with the photon index of $-2$
for the flux calculation. The simple power-law decay model is indicated
by the solid line: $t^{-1.44 \pm 0.07}$. The lines labeled
``A'', ``B'' and ``C'' indicate the intervals for which spectral studies
are made. \label{fig:GRB970828_lc}
}

\figcaption[]{
The X-ray spectrum for the period (B) shown in Fig.\ref{fig:GRB970828_spec}
fitted with the model of ``absorbed power-law'' with a Radiative
Recombination edge and Continuum (RRC). \label{fig:GRB970828_spec}
}

\clearpage
% figure 1 (Light Curve)
\begin{figure}[htb]
\begin{center}
\rotatebox{270}{\scalebox{0.5}{\includegraphics{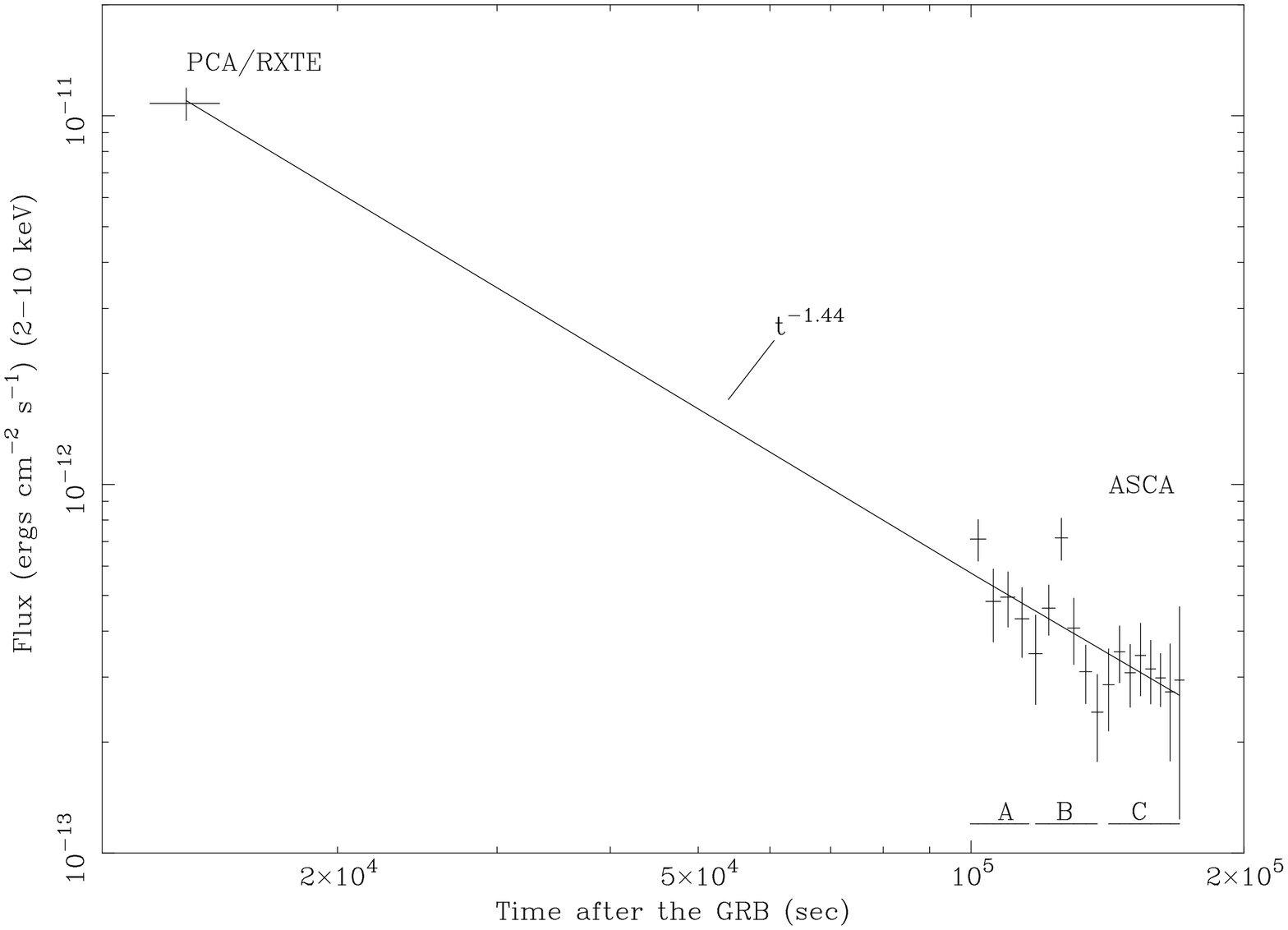}}}
\end{center}
\end{figure}

\clearpage
% figure 2 (spectrum)
\begin{figure}[htb]
\begin{center}
\rotatebox{270}{\scalebox{0.5}{\includegraphics{./fig2.ps}}}
\end{center}
\end{figure}

\clearpage
\begin{table}[htbp]
\begin{center}
\caption{Fitting Results of GRB~970828}
\label{tab:fit_res}
\vspace{2mm}
\begin{tabular}{c|c|c|c|c|c|c}\hline\hline
Period &  $N_{\rm{H}}^{\rm{gal}}$ &  $N_{\rm{H}}$ &
power-law & RRC & RRC& \\
&  ($10^{20}~{\rm cm^{-2}}$)&
($10^{20}~{\rm cm^{-2}}$) &  $\Gamma$ & $E_{\rm RRC}$ (keV)
& $kT_{e}$ (keV)& $\chi^{2}~{\rm (d.o.f)}$\\
\hline
A & $3.4$ (fix) & $< 36.5$ & $1.6^{+0.2}_{-0.3}$ & --- & --- & $34.7(48)$\\
B & $3.4$ (fix) & $< 55.4$ & $2.1^{+0.3}_{-0.3}$
& $4.76^{+0.19}_{-0.25}$
& $0.8^{+1.0}_{-0.2}$ & $70.3(72)$\\
C & $3.4$ (fix) & $70.8^{+32.1}_{-27.5}$ & $2.9^{+0.7}_{-0.4}$
& --- & --- & $94.0(84)$\\\hline
\end{tabular}
\end{center}
\end{table}

\clearpage
\begin{table}[htbp]
\begin{center}
\caption{Iron Line and RRC Flux of GRB~970828}
\label{tab:GRB970828_Fe}
\vspace{2mm}
\begin{tabular}{c|c|c}\hline\hline
{\bf period}&  Line ($6.97$ keV) &  R.R.C ($9.28$ keV) \\
& (${\rm photon~cm^{2}s^{-1}}$) & (${\rm photon~cm^{2}s^{-1}}$) \\
\hline
A & $< 4.7 \times 10^{-6}$ & $< 1.2 \times 10^{-5}$\\
B & $< 1.5 \times 10^{-6}$ & {\bf $1.7^{+6.4}_{-1.2} \times 10^{-5}$}\\
C & $< 3.0 \times 10^{-5}$ & $< 7.0 \times 10^{-6}$  \\\hline
\end{tabular}
\end{center}
\end{table}

\end{document}